\documentclass[
aps,
prb,
reprint,
superscriptaddress,
amsmath,amssymb,
longbibliography,
]{revtex4-1}

\usepackage[version=3]{mhchem} 
\usepackage{graphicx}
\usepackage{dcolumn}
\usepackage{bm}
\usepackage{natbib,hyperref}
\hypersetup{colorlinks=true, allcolors=blue}

\renewcommand{\emph}[1]{\textit{#1}}

\begin{document}

\title{Spin splitting in 2D monochalcogenide semiconductors}
\author{Dat T. Do}
\email[]{dodat@msu.edu}
\author{Subhendra D. Mahanti}
\author{Chih Wei Lai}
\email[]{cwlai@msu.edu}
\affiliation{Department of Physics and Astronomy, Michigan State University, East Lansing, MI 48824, USA}

\date{\today}

\begin{abstract}
We report \emph{ab initio} calculations of the spin splitting of the uppermost valence band (UVB) and the lowermost conduction band (LCB) in bulk and atomically thin GaS, GaSe, GaTe, and InSe. These layered monochalcogenides appear in four major polytypes depending on the stacking order, except for the monoclinic GaTe. Bulk and few-layer $\epsilon$- and $\gamma$-type, and odd-number $\beta$-type GaS, GaSe, and InSe crystals are noncentrosymmetric. The spin splittings of the UVB and the LCB near the $\Gamma$-point in the Brillouin zone are finite, but still smaller than those in a zinc-blende semiconductor such as GaAs. On the other hand, the spin splitting is zero in centrosymmetric bulk and even-number few-layer $\beta$-type GaS, GaSe, and InSe, owing to the constraint of spatial inversion symmetry. By contrast, GaTe exhibits zero spin splitting because it is centrosymmetric down to a single layer. In these monochalcogenide semiconductors, the separation of the non-degenerate conduction and valence bands from adjacent bands results in the suppression of Elliot-Yafet spin relaxation mechanism. Therefore, the electron- and hole-spin relaxation times in these systems with zero or minimal spin splittings are expected to exceed those in GaAs when the D'yakonov-Perel' spin relaxation mechanism is also suppressed.
\end{abstract}


\maketitle


\section{Introduction}

Potential applications in spin-dependent electronics and optoelectronics have driven the search for materials capable of exhibiting a high degree of spin polarization and long spin relaxation time \cite{zutic2004,dyakonov2008}. However, optical generation of electron and hole spin polarization and resulting polarized luminescence are typically limited by the mixing of degenerate valence bands in most semiconductors \cite{dyakonov2008}. Recent reports of valley polarization in atomically thin transition metal dichalcogenides (TMDs) \cite{xiao2012,cao2012,zeng2012,mak2012,xu2014} suggest potential exploitation of both spin and valley degrees of freedom for electronics and optoelectronics. In an experimental study \cite{tang2015}, we demonstrated the high generation and preservation of optical spin polarization and dynamics in a group-III monochalcogenide, GaSe, under nonresonant optical pumping. The observed near unity optical spin polarization \cite{gamarts1977,ivchenko1977} is attributed to suppressed electron and hole spin relaxation rates resulting from reduced valence-band mixing. However, the microscopic spin relaxation mechanisms in GaSe and related monochalcogenides are not fully understood.

In metals and semiconductors, the major spin relaxation mechanisms-- including Elliott-Yafet (EY) \cite{elliott1954,yafet1963} and D'yakonov-Perel' (DP) \cite{dyakonov1971,dyakonov1972,dyakonov1984} mechanisms-- are associated with the spin-orbit interaction (SOI) and the spin-orbit-induced spin splitting, $\Delta_s (\vec{k}) = |E(\vec{k}, \uparrow) - E(\vec{k}, \downarrow)|$ \cite{dyakonov2008,boross2013}. Considering spin-relaxation with a four-state (two bands with spin) model Hamiltonian in the absence of an external magnetic field, one can relate the spin relaxation rate of electrons (holes) when the Fermi energy (corresponding to Fermi vector $k_F$) is away from the conduction (valence) band edge with the following equation \cite{boross2013}: 
\begin{equation}
	\Gamma_s \sim \frac{|\Delta_s(k_F)|^2}{\Gamma_p} + \frac{\Gamma_p |L(k_F)|^2}{\Gamma_p^2+\Delta_g^2(k_F)},
\end{equation}
where $\Gamma_p = \hbar/\tau_p$ is the scattering rate of the electron/hole, with $\tau_p$ being the corresponding momentum scattering (or correlation) time, $\Delta_s(k)$ being the spin-orbit-induced spin splitting, and $L(k)$ being the SOI between the adjacent bands with energy separation $\Delta_g$. In GaSe, the $p_z$-like uppermost valence band (UVB) is well isolated from the lowermost conduction band (LCB) ($\sim$2 eV) and the adjacent $p_{x,y}$-like valence bands, and as a result $L/\Delta_g \approx$ 0.02--0.04 \cite{kuroda1980,kuroda1981}. The hole-spin relaxation due to the EY mechanism $\Gamma_s^{EY}\approx (\frac{L}{\Delta_g})^2 \Gamma_p$ is thus expected to be much smaller than the momentum relaxation rate $\Gamma_p$. The spin relaxation caused by the DP mechanism can be seen as being due to the precession of spins in an effective magnetic field associated with $\Delta_s(k)$ \cite{dyakonov1984,dyakonov2008,boross2013}. The DP spin relaxation rate is proportional to the spin splitting, $\Gamma_s^{DP} \propto \tau_p \Delta_s(k)^2$, where $\tau_p$ is the momentum relaxation time. Therefore, when the spin relaxation is dominated by the DP mechanism, the smaller the spin splitting, the longer the spin relaxation time $\tau_s = \hbar/ \Gamma_s$ for the same momentum relaxation rate $\Gamma_p$. 

 To understand the spin relaxation, one first needs the momentum($\vec{k}$)-dependent $\Delta_s(\vec{k})$ of the bands near the fundamental gap. In the absence of magnetic fields, $\Delta_{s}(\vec{k})$ is zero in centrosymmetric crystals because of the constraints of time-reversal symmetry [$E(\vec{k},\uparrow) = E(-\vec{k},\downarrow)$, Kramers degeneracy] and spatial inversion symmetry [$E(\vec{k},\uparrow) = E(-\vec{k},\uparrow)$]. When the inversion symmetry is broken in crystals (bulk inversion asymmetry (BIA)) \cite{dresselhaus1955} or heterostructures (structural inversion asymmetry (SIA)) \cite{bychkov1984,winkler2000}, $\Delta_s(\vec{k})$ is finite, and only the Kramers degeneracy is left. Understanding $\Delta_s (\vec{k})$ in GaAs and other zinc-blende semiconductors has been a subject of considerable interest \cite{yu2005,chantis2006,chantis2010,luo2009,luo2010,shen2010} since the seminal work of Dresselhaus \cite{dresselhaus1955}. \emph{Ab initio} calculations, such as LDA (or GGA) and self-consistent GW methods, of $\Delta_s(\vec{k})$ in bulk GaAs and two-dimensional GaAs-based superlattices and heterostructures have improved the understanding of the spin splitting \cite{chantis2006,chantis2010,luo2009,luo2010}. A few theoretical calculations of $\Delta_s (\vec{k})$ in TMDs also have been reported \cite{zhu2011,chang2014}. In this study, we report \emph{ab initio} calculations of $\Delta_s(\vec{k})$ of the uppermost valence band (UVB) and the lowermost conduction band (LCB) in GaSe and related group-III monochalcogenides, including GaS, GaTe, and InSe.

\begin{figure}[htb]
\centering
\includegraphics[width=0.4\textwidth]{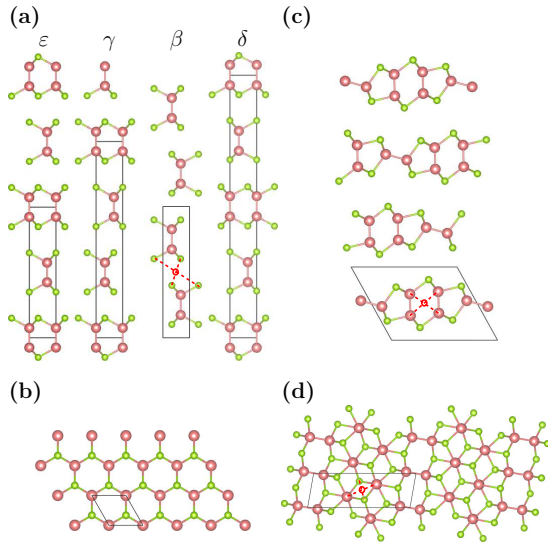}
\caption{\label{fig:xtal}
(a) Side view of the 2Ha $\epsilon$-, 3R $\gamma$-, 2Hb $\beta-$, and 4H $\delta$-polytype $MX$ ($M$ = Ga, In; $X$ = S, Se) unit cell. (b) Top view of the $MX$ single layer. (c) Side view of the monoclinic \ce{GaTe} unit cell. (d) Top view of \ce{GaTe} single layer. $M$ and $X$ are big (brown) and small (green) spheres, respectively. In the centrosymmetric systems, one possible inversion center is denoted by a red circle.
} 
\end{figure}

Monochalcogenides $MX$ ($M$ = \ce{Ga}, \ce{In}; $X$ = S, Se) crystallize in hexagonal layered structures \cite{madelung2004} (Fig.~\ref{fig:xtal}) of four major polytypes, namely $\epsilon$, $\gamma$, $\beta$, and $\delta$ (Fig.~\ref{fig:xtal}a), depending on the stacking order (hereinafter referred to as $MX$ crystals). $\epsilon$-, $\gamma$-, $\beta$-, and $\delta$-$MX$ crystals belong to the space group (Schoenflies notation) of $D_{3h}^1$, $C_{3v}^5$, $D_{6h}^4$, and $C_{6v}^4$, respectively. Monolayer $MX$ crystals (space group $D_{3h}$) are noncentrosymmetric. Bulk $\epsilon$-, $\gamma$-, and $\delta$-$MX$ crystals, which appear in an AB, ABC, and ABCD stacking order, are noncentrosymmetric, while $\beta$-$MX$ crystals are centrosymmetric with an AB stacking order. Additionally, there are two exceptions: (1) an atomically thin $\beta$-$MX$ crystal with even-number layers is centrosymmetric, and (2) a bilayer $\delta$-$MX$ crystal can be identical to either a bilayer $\epsilon$-$MX$ crystal (noncentrosymmetric) or a bilayer $\beta$-$MX$ crystal (centrosymmetric) depending on which two layers are isolated from a bulk $\delta$-$MX$ crystal. GaTe appears as a distorted form of the $MX$ structure, where one out of three Ga-Ga bonds lies in the $a$-$b$ plane (Fig.~\ref{fig:xtal}c--d). In contrast to $MX$ crystals, GaTe crystals belong to the monoclinic lattice system (space group $C_{2h}^3$), and are centrosymmetric down to a single layer \cite{madelung2004}. 

Bulk and few-layer $\epsilon$- and $\gamma$-type, as well as odd-number few-layer $\beta$-type GaS, GaSe, and InSe crystals finite spin splittings, while bulk and even-number few-layer $\beta$-type GaS, GaSe, and InSe as well as GaTe crystals exhibit zero spin splitting. The difference is due to the constraints of the aforementioned time-reversal and spatial inversion symmetry (or the lack of it).

\section{Computational Methods}
We compute the band structures and $\Delta_s(\vec{k})$ of valence and conduction bands with the projector augmented wave method as implemented in the VASP \cite{kresse1993,kresse1996,kresse1996a,kresse1999,blochl1994} package and the full-potential (linearized) augmented plane-wave as implemented in the WIEN2k \cite{scwartz2003,singh2006} package. The band structures are calculated with the WIEN2k package, with the optimized crystal structures determined by minimizing the total energy with all electrons (including core electrons) with VASP. In all calculations, exchange-correlation energies are determined by the Perdew-Burke-Ernzerhof (PBE) \cite{perdew1996} generalized gradient approximation (GGA) \cite{martin2004}, which systematically underestimates the band gaps and produces $E(\vec{k})$ dispersions (effective masses) different from experimental values. These shortcomings of the GGA also limit the accuracy of the calculated $\Delta_s(\vec{k})$.

The spin-orbit interaction (SOI) is included in our calculations of the overall band structure and the spin splitting of a given band in a self-consistent manner using a second variation approach \cite{koelling1977,macdonald1980,kleinman1980}. The SOI Hamiltonian in the spherical symmetric potential can be represented as: $H_{so} = 1/ (2 m_e^2 c^2) \times [1/r \ dV(r)/ dr] \vec{L} \cdot \vec{S}$, where $m_e$ is the electron mass, c the speed of light, $\vec{L}$ and $\vec{S}$ the orbital and spin momentum vectors, and $V(r)$ an effective single particle local potential seen by the electron. This form of $H_{so}$ is correct as long as $V(r)$ is local and isotropic. In Hartree approximation and LDA, the effective potential is indeed local, though it is not always isotropic. The isotropic approximation is valid because the dominant contribution to $H_{so}$ is from regions near the nucleus. However, local approximations do not give correct band structure near the band gap. The accuracy of the band gap can be improved with hybrid models such as the Heyd-Scuseria-Ernzerhof (HSE06) \cite{heyd2003} (a mixture of non-local and local exchange) or GW-like theories. In these approximations, the aforementioned simple form of $H_{so}$ is not necessarily valid \cite{blume1962a,blume1962b}. A compromised approach invloves  calculating the band structure using non-local or GW theories while calculating the spin splitting with a local form of the $H_{so}$ and the LDA potential \cite{chantis2006,sakuma2011}. Nonetheless, one has to examine uncertainties in these hybrid calculations critically.

To model a few-layer thin film, we create a supercell (supercell method \cite{martin2004}) containing one few-layer structure and a 15--25 \AA{} thick vacuum spacer, which is large enough to suppress interactions arising from the artificial periodicity present in the supercell method. The crystalline $c$-axis of the supercell is set perpendicular to the crystalline $a$-$b$ plane. In this way, one can distinguish the effects of intra- and inter-layer interactions on the electronic structures in few-layer structures. The number of atoms in a unit cell is as follows: eight for $\epsilon$- and $\beta$-$MX$, twelve for $\gamma$-$MX$, sixteen for $\delta$-$MX$, and twelve for monoclinic GaTe. To obtain an energy accuracy of 0.1 meV in self-consistent calculations, we use $\Gamma$-centered Monkhorst--Pack \cite{monkhorst1976} $\vec{k}$-meshes of $24\times24\times4$ and $24\times24\times1$ for bulk and few-layer GaSe-type structures, respectively. For GaTe, we use meshes of $16\times6\times8$ and $16\times6\times2$ for bulk and few-layer GaTe, respectively.

\section{Results}

\subsection{Band structure: bulk versus single-layer}

\begin{figure}[htb]
\centering
\includegraphics[width=0.45\textwidth]{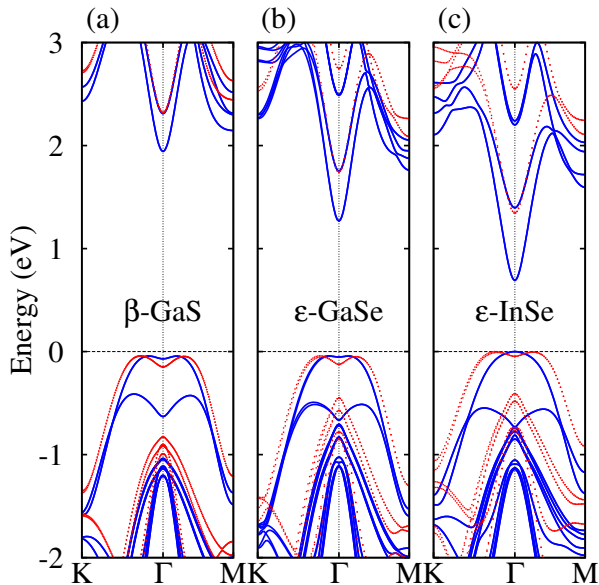}
\caption{\label{fig:bandstructure}
Electronic band structures, along K--$\Gamma$--M in the hexagonal Brillouin zone (BZ), of bulk (solid blue curves) and monolayer (dotted red curves) (a) $\beta$-\ce{GaS}, (b) $\epsilon$-\ce{GaSe}, and (c) $\epsilon$-\ce{InSe}. The zero energy is set at the valence band maximum.
}
\end{figure}

In Fig.~\ref{fig:bandstructure}, we show the electronic band structures of bulk and monolayer $\beta$-GaS, $\epsilon$-GaSe, and $\epsilon$-InSe, which are the most naturally abundant. The general features of the electronic band structures, except the spin splitting, are nearly polytype-independent, owing to the weak inter-layer interactions. The lowermost conduction band (LCB) has $s$-like symmetry, whereas the two uppermost valence bands (UVBs) have $p_z$-like symmetry. The $p_{x,y}$-like valence bands appear $\sim$1 eV below the UVB as a result of the crystal field and SOI. The calculated band structures for $\epsilon$-GaSe show a nearly direct band gap at the $\Gamma$-point of the Brillouin zone (BZ), where a valley appears in the UVB. The energy of the LCB at the $\Gamma$-point is $\sim$0.5 eV lower than that at the M point, consistent with the hybrid density functional calculations \cite{zolyomi2013}. On the contrary, tight binding calculations  show that the energy of the LCB at the M point for GaSe is $\sim$10 meV below that at the $\Gamma$-point in the BZ \cite{camara2002}.

The band gap is seen to decrease with increasing atomic number (Ga$\rightarrow$In or S$\rightarrow$Se). The calculated band gaps are 2.0 eV, 1.3 eV, and 0.71 eV for $\beta$-GaS, $\epsilon$-GaSe, and $\epsilon$-InSe, respectively, which are each smaller than the experimental values ($\sim$3.1 eV, 2.0 eV, and 1.3 eV) \cite{madelung2004}. The band-gap underestimation can be remedied with, for example, the HSE06 hybrid functional \cite{heyd2003,zolyomi2013,zhuang2013}. In the absence of SOI, the $p_{x,y}$ states are doubly degenerate at the $\Gamma$-point. On the other hand, the SOI lifts this energy degeneracy with a spin-orbit splitting $\Delta_{SO} \approx$ 0.09 eV, 0.34 eV, and 0.31 eV in GaS, GaSe, and InSe, respectively. $\Delta_{SO}$ in GaSe and InSe are similar in magnitude, but a factor of three smaller in GaS, agreeing with previously reported calculations \cite{mooser1973,schluter1976,doni1979,depeursinge1981}. $\Delta_{SO}$ in GaS is minimal, as expected from the weak SOI in the lighter S anions which govern the characteristics of the few uppermost valence bands of GaS. Monolayer GaS, GaSe, and InSe have very similar band structures (Fig.~\ref{fig:bandstructure}). We note two different features in the band structures of monolayer $MX$s in comparison with their bulk counterparts: (1) the quantum confinement along the $c$-axis increases the band gap to 2.36 eV, 1.78 eV, and 1.4 eV for GaS, GaSe and InSe, respectively, and (2) the band gap becomes indirect as the valley at the $\Gamma$-point becomes wider in momentum ($k = |\vec{k}|$) and deeper in energy ($E$). 

\begin{figure}[htb]
\centering
\includegraphics[width=0.45\textwidth]{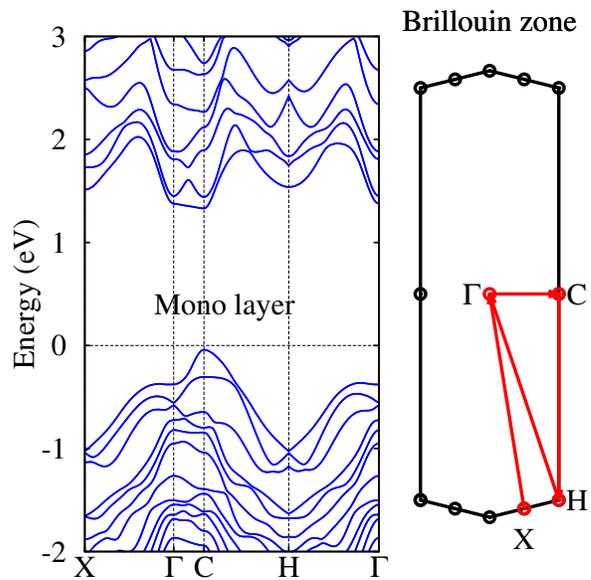}
\caption{\label{fig:GaTe_bandstructure}
The electronic band structure of monolayer \ce{GaTe} (left) along the selected high-symmetry directions in the 2D BZ (right).
}
\end{figure}

The band structure of GaTe (bulk) has also been calculated with GGA \cite{sanchez-royo2002,rak2009b}, showing a direct band gap of $\sim$1 eV. The inclusion of SOI causes negligible changes in the UVB and LCB of GaTe. Monolayer GaTe shows a direct band gap of 1.4 eV (Fig.~\ref{fig:GaTe_bandstructure}), with LCB having two nearly degenerate minima at the $\Gamma$ and C points. At the C point, LCB has $s$-like symmetry while the UVB has $p_y$-like symmetry. SOI removes the  $p_{x,y}$ degeneracy of valence bands at $\Gamma$, with $\Delta_{SO}\approx 0.2$ eV. $\Delta_{SO}$ is smaller in GaTe than in GaSe despite Te being heavier than Se. The reduction in the strength of $\Delta_{SO}$ is due to the quenching of orbital angular momentum in the lower symmetry crystalline structure, as demonstrated by a sizable $\Delta_{SO} \approx$ 0.7 eV calculated for a hypothetical $\beta$-type GaTe (space group $D_{6h}^4$). 

\subsection{Spin splitting}
\begin{figure}[htb]
\centering
\includegraphics[width=0.45\textwidth]{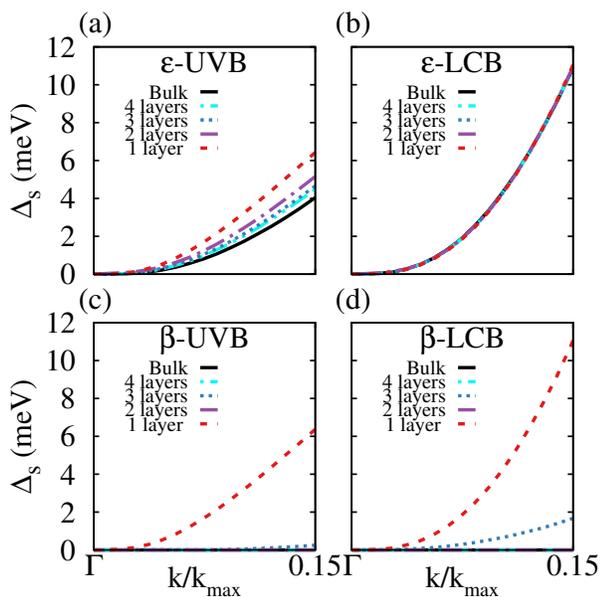}
\caption{\label{fig:GaSe_spin_splitting}
Spin splitting $\Delta_s(k')$ as a function of $k' = k/k_{max}$ ($k_{max}=\overline{\Gamma\text{K}}$) for the uppermost valence band (UVB) and the lowermost conduction band (LCB) along the $\Gamma$--K in $n$-layer $\epsilon$-\ce{GaSe} (a--b) and $\beta$-\ce{GaSe} (c--d) ($n$=1,2,3, and $\infty$ (bulk)).
}
\end{figure}

In Fig.~\ref{fig:GaSe_spin_splitting}, we show the spin splittings of the UVB ($\Delta_s^v(\vec{k})$) and the LCB ($\Delta_s^c(\vec{k})$) along the $\Gamma$--K direction in $\epsilon$- and $\beta$-GaSe. The spin splitting along the $\Gamma$--M direction is zero, obeying the constraint of spatial inversion symmetry. Both $\Delta_s^v(k')$ and $\Delta_s^c(k')$ decrease with the number of layers, approaching those in the bulk. At $k' = k/k_{max}= 0.15$ ($k_{max}$ is $k$ at the K point in the BZ), $\Delta_s^v \approx$ 6 meV and 4 meV for monolayer and bulk $\epsilon$-GaSe, respectively. The nearly layer-independent LCB spin splitting has a value $\Delta_s^c\approx11$ meV at $k' = 0.15$, which is slightly larger than $\Delta_s^v$.
 
In contrast to $\epsilon$-GaSe, bulk and even-number few-layer $\beta$-\ce{GaSe} crystals have zero spin splitting (Fig.~\ref{fig:GaSe_spin_splitting}c--d), obeying the constraint of spatial inversion symmetry. The $\Delta_s^v(k')$ and $\Delta_s^c(k')$ in odd-number few-layer $\beta$-GaSe crystals are finite, but diminish rapidly with increasing layers. In trilayer $\beta$-GaSe, the UVB spin splitting is less than 1 meV, and LCB spin splitting is smaller by a factor of five compared to that of the monolayer. The thickness dependent spin splitting in $\beta$-GaSe presented here are consistent with those reported in \ce{MoS2} \cite{chang2014}, which has the same symmetry as $\beta$-GaSe. Bulk $\gamma$-GaSe has similar spin splittings as bulk $\epsilon$-GaSe, with decreasing spin splittings as the number of layers increase. 

\begin{figure}[htb]
\centering
\includegraphics[width=0.45\textwidth]{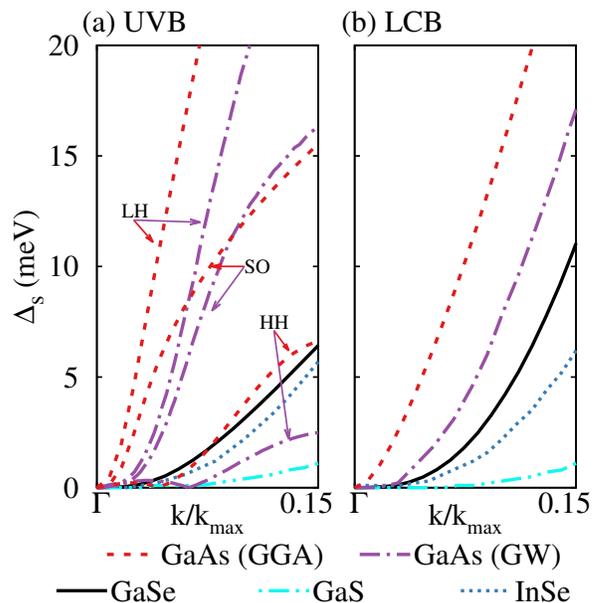}
\caption{\label{fig:compare_soi}
Spin splitting $\Delta_s(k')$ along the $\Gamma$--K line for (a) the uppermost valence band (UVB) and (b) lowermost conduction band (LCB) of monolayer GaSe, GaS, InSe, and bulk GaAs. For valence bands of GaAs, we show the splitting for the heavy hole (HH), which is the UVB, the light hole (LH) and the split-off (SO) bands. The spin splittings calculated with the GW method are extracted from Refs. \cite{chantis2006,luo2010}.
}
\end{figure}

In Fig.~\ref{fig:compare_soi}, we compare $\Delta_s^v(k')$ and $\Delta_s^c(k')$ in monolayer GaS, GaSe, and InSe (group-III monochalcogenides)  and bulk GaAs (a representative zinc-blende III-V semiconductor). Among the monolayer group-III monochalcogenides, overall spin splittings decrease from GaSe, to InSe, and then to GaS. The spin splittings typically increase with the increasing atomic number of constituent atoms as result of the enhanced SOI in the heavier atoms. However, other details of the band structure such as the band gap also contribute to the spin splittings.

The valence band of GaAs consists of a heavy hole (HH), a light hole (LH), and a split-off (SO) band \cite{yu2005,shen2010}. The calculated HH spin splitting is close to that in the UVB of GaSe. However, the spin splittings in the LH and SO bands are at least a factor of two larger than that in the UVB of GaSe. The calculated overall LCB spin splitting in GaAs is also larger than that in GaSe. The magnitude at $k'$ = 0.15 is more than two times larger in GaAs than in GaSe. The spin splitting of the heavy-hole band is reduced by about a factor of two when the GW method is used in lieu of the GGA method.

\section{Discussion}
\begin{table}[htp!]
\centering
\begin{tabular}{|l|l|l||l|l|}
\hline
Band&\multicolumn{2}{c||}{UVB}&\multicolumn{2}{c|}{LCB}\\
\hline
Coefficient&$A$ (meV)&$B$ (eV)&$A$ (meV)&$B$ (eV)\\
\hline
Monolayer&1.0&4.9&0.3&4.7\\
\hline
2-layer&2.9&2.6&2.0&4.2\\
\hline
3-layer&0.3&3.0&0.4&4.8\\
\hline
4-layer&0.6&2.6&0.4&4.9\\
\hline
5-layer&0.3&2.5&0.4&4.9\\
\hline
6-layer&0.4&2.4&0.4&4.9\\
\hline
Bulk&0.1&2.2&0.2&5.0\\
\hline
\end{tabular}
\caption{\label{tab:ABfitting} Linear ($A$) and cubic ($B$) coefficients of the $k$-dependence of spin splitting, $\Delta_s (k') =A k'+B k'^3$ with $k'=k/k_{max} \le 0.05$.
}
\end{table}

The spin splittings discussed above concern mainly the overall spin splitting up to $k'$ = 0.15. To understand the spin relaxation mechanisms, we need to identify the $k$-dependence of the spin splitting in the vicinity of the $\Gamma$-point. At small $k$, the $\mathbf{k}\cdot\mathbf{p}$ theory predicts that, in noncentrosymmetric zinc-blende and wurtzite structures, the $k$-dependence of the spin splitting contains both a linear and a cubic term when the core levels are considered \cite{dresselhaus1955,cardona1986,luo2010,voon1996}. To illustrate the $k$-dependence of the spin splitting, we fit the calculated $\Delta_s(k')$ in $\epsilon$-GaSe with the function $\Delta_s (k') = A \ k' + B \ k'^3$ for $k' <$  0.05 (Table~\ref{tab:ABfitting}). The energy scales for the coefficients $A$ (meV) and $B$ (eV) are consistent with those determined from GW calculations for GaAs \cite{chantis2006,luo2010}. Although $B$ is three to four orders of magnitude larger than $A$, there exists a crossover value of $k' = \sqrt{A/B} \sim 10^{-2}$ below which the linear term dominates. In contrast to the GaAs case where the linear term is negligible for the LCB, we find a sizable linear term for the LCB in GaSe. The cubic coefficient $B$ for the LCB is $\sim$4--5 eV for the monolayer to the bulk, with the bilayer case being slightly different. In contrast to the UVB, there appears to be an odd-even-layer effect: $B$ values for the odd layers (1 and 3) are larger, but are close to the bulk values for the even layers (2 and 4). For the LCB, the $A$ values are similar for all the layers, except for the bilayer ($A$ = 2.0 meV) and the bulk ($A$ = 0.2 meV). Note that bilayer GaSe has an unusually large $A$ value, and for the UVB, there appears to be an odd-even effect like that in the LCB. The $A$ value for the bilayer is nearly three times that for the monolayer, whereas $A$ for the four-layer is two times that for the trilayer. As pointed out in the case of GaAs, these subtle differences are due to the characteristics of the UVB and LCB energy values and wave functions, and their mixing with other bands including the core levels \cite{dresselhaus1955,cardona1986,luo2010}.

The DFT-based theories such as LDA (or GGA) underestimate band gaps and do not give accurate effective masses, resulting in overestimated $\Delta_s(\vec{k})$ \cite{chantis2006,luo2009,luo2010}. GW calculations reproduce more accurate band parameters, such as the band gap and effective mass, but are computationally more intensive than LDA (GGA) calculations. For simplicity, in this work, we have used GGA to calculate $\Delta_s(\vec{k})$. The GGA calculation underestimates the GaAs band gap by a factor of ten; however, the spin splitting only deviates from that determined by the GW calculation by a factor of two. The band gaps are underestimated by the GGA for GaSe and related monochalcogenides by a factor of approximately two, which is significantly less than that for GaAs. For example, in GaSe, the GGA calculation gives a band gap of about 1 eV, which is off from the GW/HSE06 calculation \cite{zhuang2013} and the measured band gap ($\sim$2 eV) \cite{mooser1973} by a factor of two. Therefore, we expect the GGA calculation to produce spin splittings  close to the value obtained with the GW calculation. We also expect similar variations of $\Delta_s$ with $k$ from one conduction/valence band to another and from bulk to atomically thin layers.

\section{Conclusion}
We present a systematic study of spin-orbit-induced spin splittings bulk and atomically thin group-III monochalcogenides $MX'$ ($M$ = Ga, In; $X'$ = S, Se, Te). The spin splitting vary with anion element and crystal symmetry. Centrosymmetric crystals, including bulk $\beta$-type GaS, GaSe, and InSe, as well as monoclinic GaTe down to the monolayer, have zero spin splitting, as anticipated from the constraints of spatial inversion symmetry and time-reversal symmetry. In these monochalcogenide semiconductors, the separation of the non-degenerate conduction and valence bands from other adjacent bands results in suppression of Elliot-Yafet spin relaxation mechanism. Therefore, the electron and hole spin relaxation times in these systems with zero or minimal spin splittings and reduced valence-band mixing are expected to be longer than those in a zinc-blende semiconductor (eg., GaAs \cite{hilton2002,yu2005,shen2010}), owing to the suppression of D'yakonov-Perel' and Elliot-Yafet spin relaxation mechanisms.

\begin{acknowledgments}
This work was supported by NSF grant DMR-09055944 as well as a start-up funding and the J.~Cowen endowment at Michigan State University. The calculations were performed with computational resource provided by Institution for Cyber Enabled Research (ICER) and High Performance Computer Center (HPCC) at Michigan State University.
\end{acknowledgments}

\bibliography{/Users/cwlai/Dropbox/Bib/lai_lib}
\end{document}